\documentclass{kapproc}

\usepackage{procps-mac} 

\usepackage[dvips]{graphicx}
\input epsf
\input{epsf.sty}
\upperandlowercase

 \normallatexbib %will produce bibliography entries as shown in the
\begin{document}

\articletitle[Disorder in Underdoped Bi-2212 Superconductor]{First Order Transition of the Vortex Lattice in Disordered 
BI-2212 Crystals}

%\articlesubtitle{This is an Article Subtitle}

\author{Kees van der Beek, Irina Abalosheva, and Marcin Konczykowski}
\affil{Laboratoire des Solides Irradi\'{e}s, CNRS UMR - 7642 \& CEA / DSM / DRECAM\\ Ecole Polytechnique, F-91128 Palaiseau cedex, France}
%\email{beek@polytechnique.fr}

\author{Ming Li and Peter Kes}
\affil{Kamerlingh Onnes Laboratorium der Rijksuniversiteit Leiden\\ P.O. Box 9506, NL - 2300 RA Leiden, the Netherlands}
%\email{kes@Phys.LeidenUniv.nl}

\author{Mikhail Indenbom}
\affil{Laboratoire de Magn\'{e}tisme de Bretagne, Universit\'{e} de Bretagne Occidentale \\ 6, Avenue Le Gorgeu, F-29185 Brest, France}
%\email{indenbom@issp.ac.ru}

\begin{abstract}
Using differential magneto-optical imaging, we address the question of 
mesoscopic inhomogeneity in underdoped 
Bi$_{2}$Sr$_{2}$CaCu$_{2}$O$_{8}$ (Bi-2212) single crystals.
Among other features, it is shown that an anomalous temperature dependence of 
the penetration field and of the first order transition (FOT) field 
of the vortex lattice in such crystals can be understood as arising from 
inhomogeneity. The effect of chemical inhomogeneity and pinning on flux 
penetration and the FOT is discussed. 
\end{abstract}


\section{Introduction}
It is commonly assumed that the vortex lattice undergoes a first 
order transition (FOT) to the so-called vortex-liquid state in very clean, 
defect free superconducting single crystals 
only \cite{Safar92II,Kwok92,Zeldov95II}. In disordered crystals, the 
transition to a vortex liquid is presumed to be second order, 
although, at present, only very few reports substantiate this 
\cite{Klein96,Okuma2001}. In the high temperature superconductors, 
the FOT is observed irrespectively of the oxygen content of the material
%Since the latter also determines the superfluid density and the material 
%anisotropy factor $\varepsilon \ll 1$, variation of the oxygen content has 
%revealed a ``universal'' dependence on $\varepsilon$ of the 
%induction at which the FOT takes place, $B_{FOT} \propto \varepsilon^{-2}$
\cite{Sasagawa98}
. 

The FOT in oxygen-underdoped 
Bi$_{2}$Sr$_{2}$CaCu$_{2}$O$_{8}$ (Bi-2212) deserves special attention. 
Single crystals of this material cannot be obtained by annealing 
optimally doped material, and only recently has a technique 
been developped to grow them \cite{MingLi2002I}. Even 
then, underdoped Bi-2212 crystals present nanometer-scale inhomogeneity of the  
superconducting gap \cite{Pan,Hoogenboom}, 
as well as chemical inhomogeneity on larger scales \cite{Noriko}.

Here, we present Hall-probe array magnetometry and 
differential mag-neto-optical (DMO) \cite{Soibel2000} imaging experiments 
on strongly underdoped Bi-2212 single crystals. The 
local induction measured by both techniques shows a well-resolved 
discontinuity at the same FOT field, $B_{FOT}$. DMO allows us to map 
the spatial variation of $B_{FOT}$ within a crystal, and to obtain its 
temperature dependence in different locations. It is argued that the 
spatial variation of $B_{FOT}$ is responsible for magnetization 
anomalies near the critical temperature, $T_{c}$.

\section{Experimental Details}

\subsection{Sample growth}

Two crystalline boules of underdoped Bi$_{2}$Sr$_{2}$CaCu$_{2}$O$_{8}$  were 
grown using the travelling-solvent floating zone method,
under 25 mbar oxygen partial pressure. 
They were subsequently annealed at 700$^{\circ}$C in flowing N$_{2}$ gas for 
a period of one week \cite{MingLi2002I}. Crystals were separated from the boule with a razor blade.
A polarized light microscope was used to select specimens 
without obvious growth defects such as low-angle grain boundaries or 
colony boundaries. 

\subsection{Direct Magneto-optical imaging (MOI)}

The crystals were further selected using ``direct'' magneto-optical 
imaging (MOI). A crystal is covered with a Lu-iron garnet thick film 
with in--plane magnetic anisotropy, covered by an Al mirror layer
(Magistr inc., Russia), that serves as a magneto-optical indicator of 
the local induction \cite{Dorosinskii92}. After cooling down to the measurement 
temperature, a magnetic field is applied perpendicularly to the 
crystal surface and to the garnet indicator film. The local magnetic 
induction is measured by the Faraday rotation of linearly polarised 
light that twice crosses the indicator film (before and after 
reflection by the Al mirror). Bright regions in the images correspond 
to large Faraday rotation and thus a high magnetic induction, whereas 
dark areas correspond to small (or zero) Faraday rotation and 
 to a  small (or zero) local induction.

%% Figure 1 : example Bending Reference
\begin{figure}[t]
\centerline{\epsfxsize 12cm \epsfbox{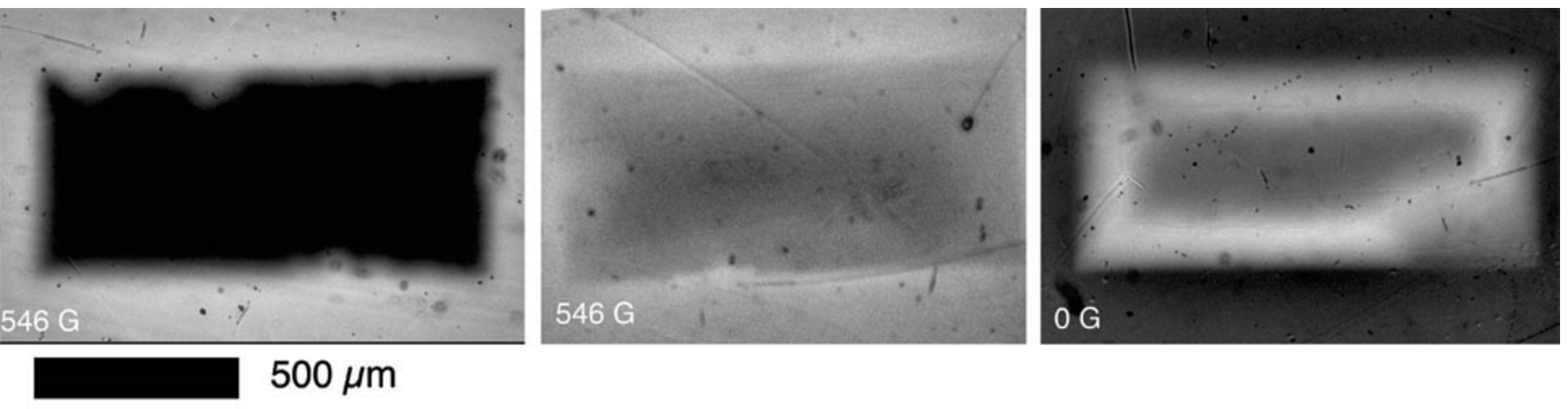}}
\centerline{\epsfxsize 12cm \epsfbox{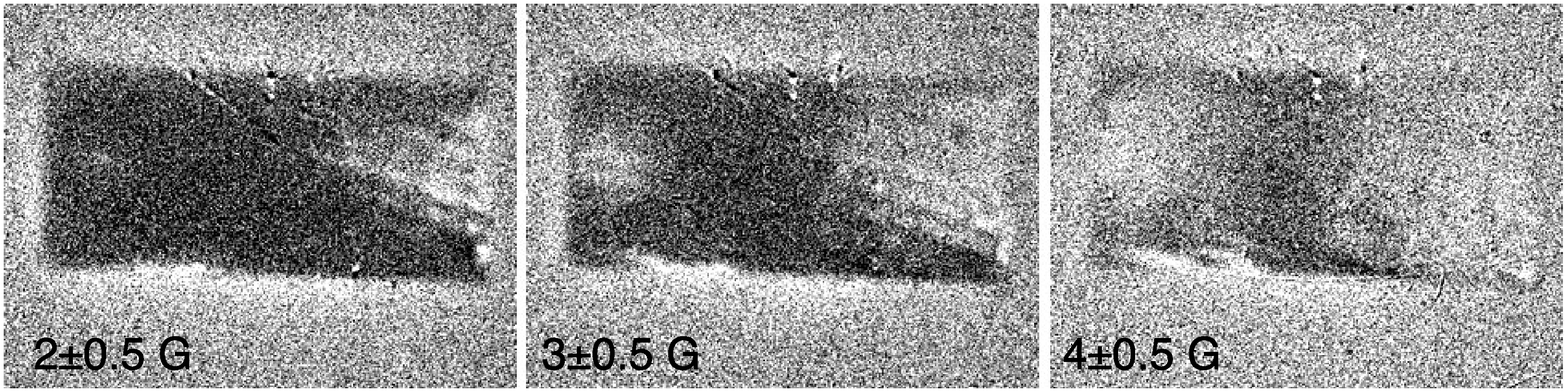}}
%\resizebox{\textwidth}{!}{\includegraphics{Bending-ref-a.eps}}
%\resizebox{\textwidth}{!}{\includegraphics{Bending-ref-b.eps}}
{\caption{(a) The top three MO images show the penetration of magnetic flux 
in underdoped  Bi$_{2}$Sr$_{2}$CaCu$_{2}$O$_{8}$ crystal A, with nominal 
$T_{c} = 65$ K, at $T = 21$ K. Left: immediately after 
the application of the external field $H = 546$ G; center: after 60 
s, magnetic flux has relaxed to the center; right: trapped flux after 
removal of the applied field.(b) The bottom three images are DMO images 
(with $\delta H = 0.5$ Oe) of flux penetration and the FOT process 
in the same crystal at 61 K. }}
\label{fig:BendingRef}
\end{figure}

Guided by direct MOI, apparently defect-free pieces were cut from larger 
crystals using a wire saw. An example of flux penetration into crystal 
A, selected by direct MO imaging from the boule 1 material 
(nominal $T_{c} = 65$ K), is shown in the top three panels of 
Figure \ref{fig:BendingRef}(a). The crystal shows homogeneous, regular 
flux penetration, that is well described by the critical state 
model \cite{Brandt96}. Judging from these images, this crystal would appear 
suitable for use in further physical measurements.

\subsection{Differential magneto-optical imaging (DMO)}

Crystals were further investigated using the DMO technique with field
modulation, $\delta H = 0.5$ Oe. All measurements were conducted after zero-field 
cooling. A field $H_{a} + \delta H$ was applied, and ten 
MO images of the flux distribution above the sample were acquired and 
summed. The applied field was then reduced to $H_{a}$, whence ten 
other images were acquired and successively subtracted from the first 
sum. This procedure was repeated twenty times; the resulting twenty 
differential images were averaged to produce the final DMO images. 
These images should be interpreted as representing a map 
of the local ``permeability'' of the sample. The grey level outside the 
sample boundaries represent a ``permeability'' $\Delta B / \Delta H = 1$. 
A zero intensity (black) represents diamagnetic screening, {\em i.e.} 
$\Delta B / \Delta H = 0$, while intermediate grey levels represent 
partial screening of the field modulation. Clear regions indicate a 
paramagnetic response. This is of particular interest near
the vortex lattice FOT: the discontinuity in the local induction at the 
transition \cite{Zeldov95II} results in 
a change $\Delta B$ that is larger than $\Delta H$ by an amount 
$\Delta B_{FOT}$, and thus in a 
paramagnetic signal \cite{Soibel2000}. DMO images of the FOT process at 61 K in crystal A are shown in the bottom panels of 
Fig.~\ref{fig:BendingRef}. The FOT first takes place on the line
defects oriented at $\sim 20^{\circ}$ with respect to the long 
crystal side \cite{Kes2003}, located near the crystal edge, and 
progressively moves inwards to the crystal center. Regions of the crystal
that have not yet undergone the FOT fully screen the field modulation.

\begin{figure}[t]
    %\resizebox{\textwidth}{!}{\includegraphics{Fig-2.pdf}}
    \centerline{(a)\epsfxsize 8cm \epsfbox{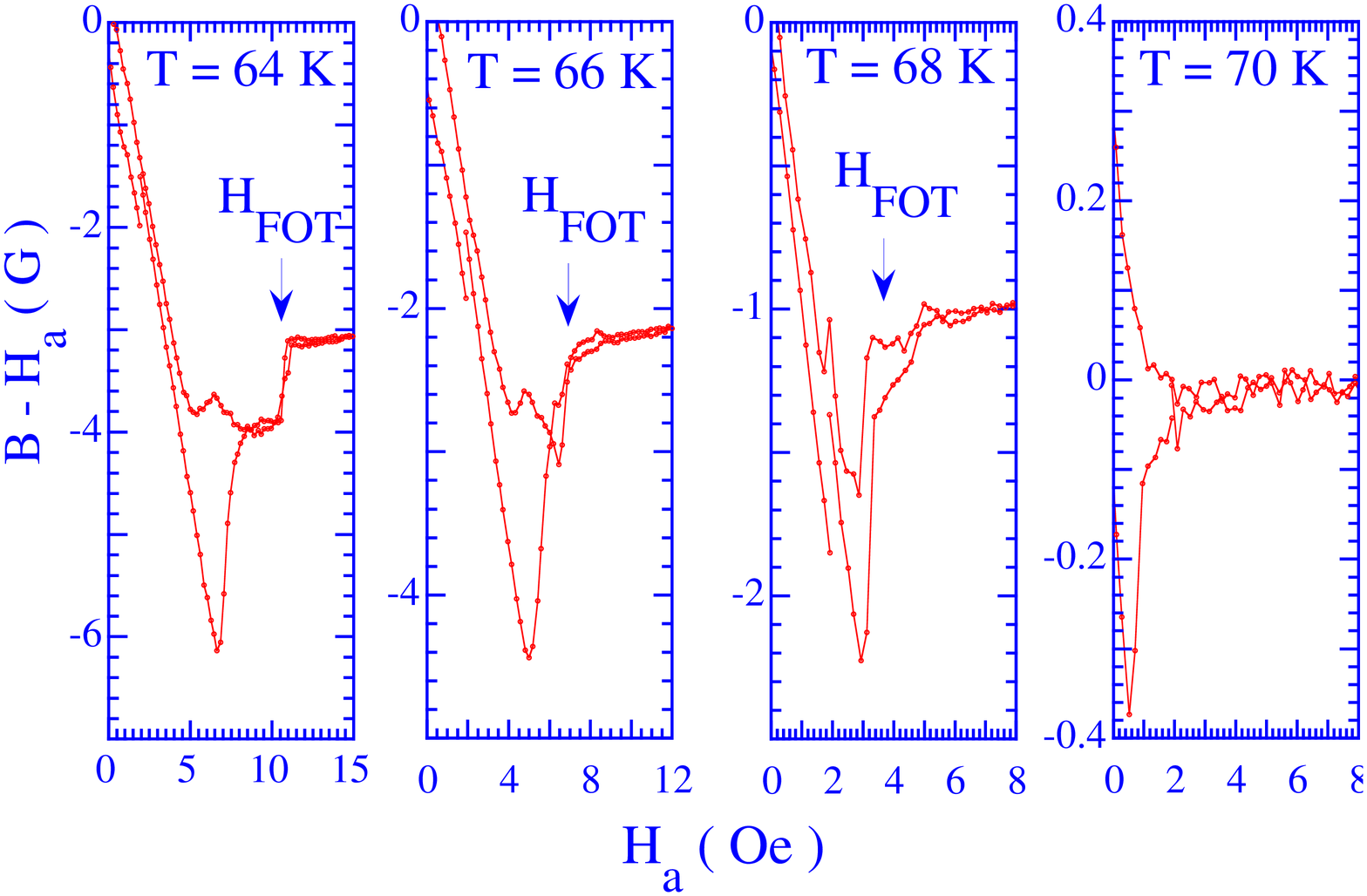}(b)\epsfxsize 2.4cm 
    \epsfbox{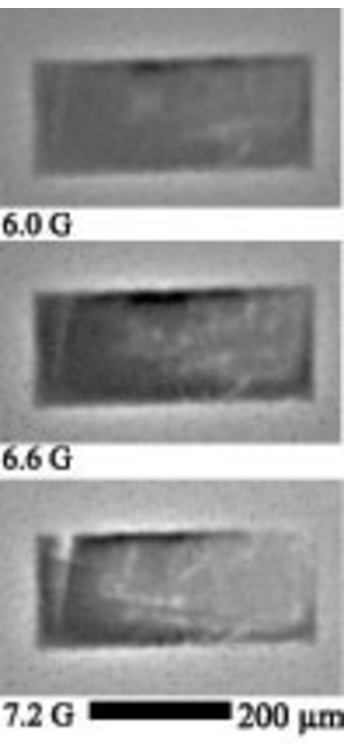}}
    \caption{(a) Loops of the local ``magnetisation'' $B - H_{a}$ 
    measured near the center of crystal B (nominal $T_{c} = 75$ K) 
    using an array of micoscopic Hall sensors. (b) DMO images of the FOT 
    process on the surface of the same crystal, at $T = 66.2 $ K.}
    \label{fig:Hall}
\end{figure}

\subsection{Local Hall probe magnetometry}

Flux distributions over a number of crystals selected by direct MOI were measured using Hall 
probe magnetometry. A crystal is placed on an array of 11 
Hall probes, of dimensions $10 \times 10$ $\mu$m$^{2}$, located 10 $\mu$m 
apart \cite{Zeldov95II}, with the array parallel to the shorter 
crystal edge, and traversing the crystal center. The local induction at each  
sensor is measured as the applied field $H_{a}$ is cycled, at constant 
temperature, from zero to a maximum field, and back to zero. Typical 
loops of the local ``magnetisation'', $B - H_{a}$, measured at a 
probe near the center of crystal B (originating from boule 2 with nominal $T_{c} = 75$ K) 
are shown in Fig.~\ref{fig:Hall}(a).

% Figure 3 : phase diagram
\begin{figure}
    %\resizebox{\textwidth}{!}{\includegraphics{UD-phase-diagram.pdf}}
    \centerline{\epsfxsize 10cm \epsfbox{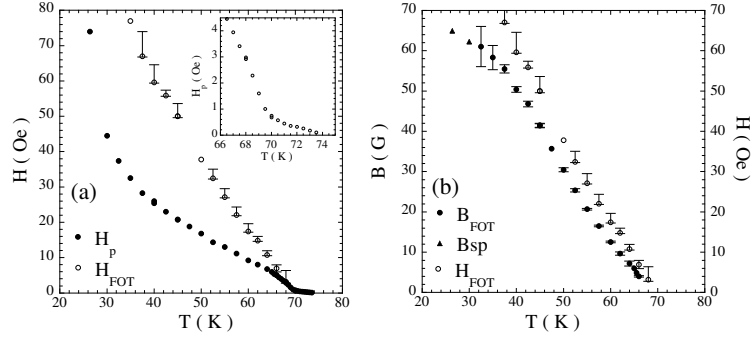}}
        \caption{(a) Field $H_{p}$ of flux penetration to the center of crystal 
    B, and field $H_{FOT}$ at which the FOT takes place there; (b) local induction at which
    the FOT takes place. Error bars show spread of $H_{FOT}$ and 
    $B_{FOT}$ in the crystal. The inset to (a) shows an enlarged view
    of $H_{p}(T)$ near $T_{c} \sim 75$ K. }
    \label{fig:phase diagram}
\end{figure}

\section{Phase diagram and FOT process}

From the cusp in the loops of local induction (Fig.~\ref{fig:Hall}), 
the field of flux penetration to the Hall probe 
position is identified, {\em e.g.} $H_{p}(64$ K) = 6.9 Oe. The step $\Delta B_{FOT}$ in $B - 
H_{a}$ signals the field at which the FOT occurs, {\em e.g.} $H_{FOT}(64$ K) = 11 Oe. 
DMO imaging on crystal B at $T = 66$ K shows that the FOT process starts
independently, at $H_{a} = 6.6$ Oe, in different spots scattered throughout 
the crystal. Such inhomogeneity is found both with the Hall array 
and with DMO. For  example, at $H_{a} = 7.2$ Oe the central area 
of the crystal is already in the vortex liquid state, but the 
vortex lattice near the two line intergrowths \cite{Kes2003} at the
bottom transits to the liquid only here.

Figure~\ref{fig:phase diagram} renders $H_{p}$, $H_{FOT}$, and $B_{FOT}$, measured 
over crystal B, as function of temperature. All three decrease 
gradually as function of $T$, up to $T = 66$ K, at which $H_{p}(T)$ starts to 
decrease more rapidly. Such a ``collapse'' of the 
penetration field was previously reported by Mrowka {\em et al.} 
\cite{Mrowka99}, but they did not report on any features 
associated with the FOT. Note that in the temperature range of 
the ``collapse'', 66$ < T <$ 70 K, the step in $(B - H_{a})$ signaling the
FOT cannot be discriminated from the rapid increase of $B$ at $H_{p}$ 
[Fig.~\ref{fig:Hall}(a)]. Above 70 K, $H_{p}$ is reduced to less than 0.6 Oe
and the FOT can no longer be observed at all, be it in Hall array measurements
or in DMO. The inset of 
Fig.~\ref{fig:phase diagram} shows that superconductivity in crystal B
persists up to $T \approx 75$ K: $H_{p}$ slowly decreases from 
$H_{p}(70 \,{\rm K}) = 0.6$ G to 0.

% Figure 5 : phase diagram crystal 1-5 measured by Irina Abalosheva
\begin{figure}
    %\resizebox{\textwidth}{!}{\includegraphics{UD-phase-diagram.pdf}}
    \centerline{(a) \epsfxsize 5cm \epsfbox{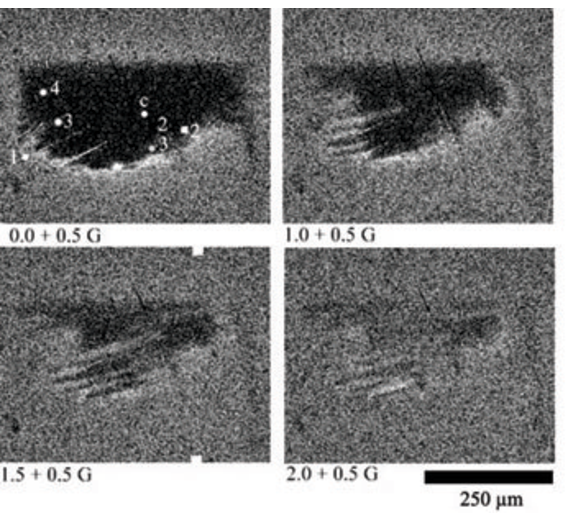} (b) \epsfxsize 
    6cm \epsfbox{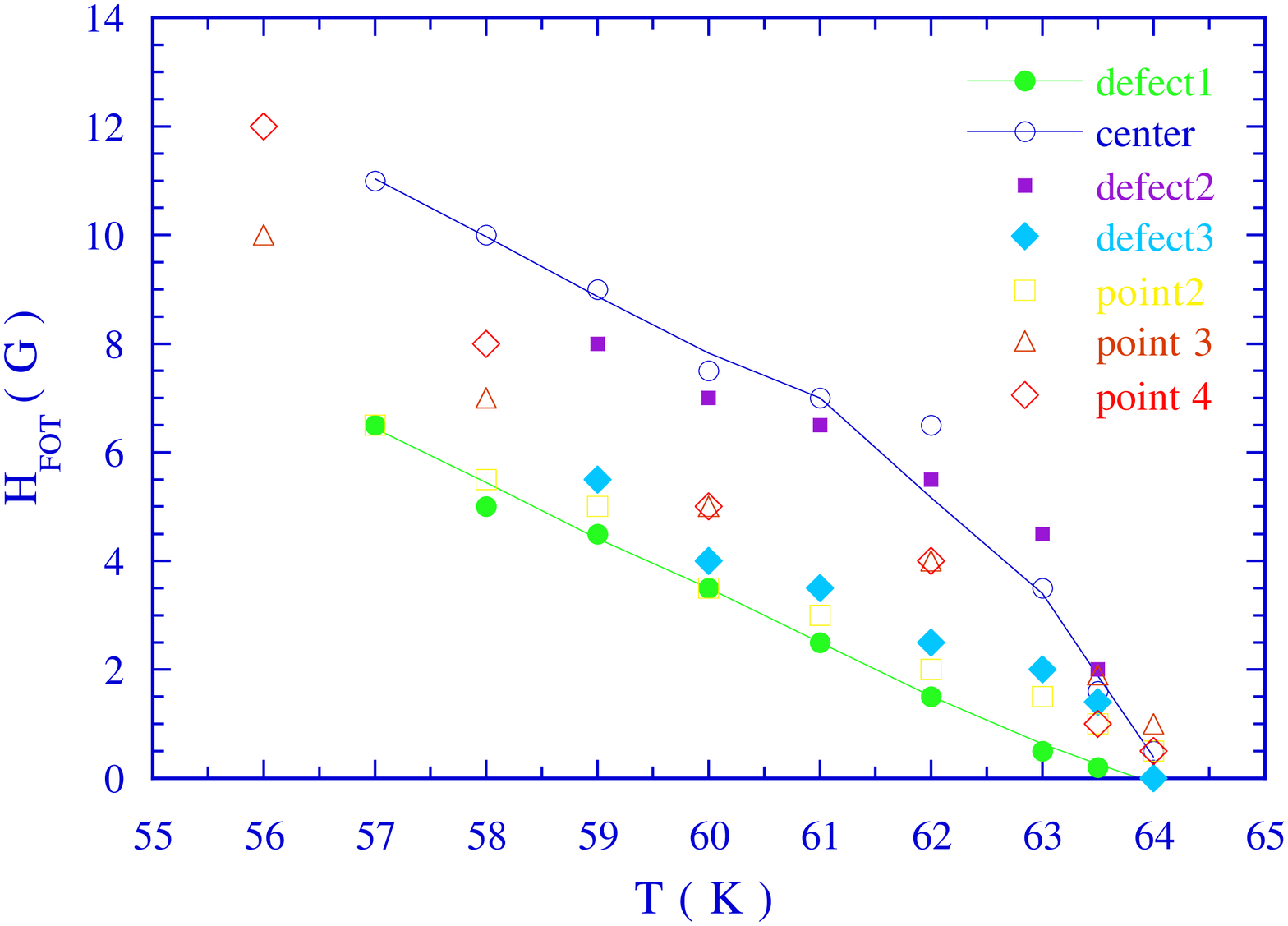}}
    \caption{(a) DMO images of the FOT process in crystal C at 63.5 K.
    The FOT propagates inwards from the crystal edge, 
    along crystalline defects. (b) $B_{FOT}$ as 
    function of temperature, measured in the different locations 
    indicated in the first frame of (a). Open symbols denote 
    apparently ``homogeneous'' regions 1-4 (including the crystal center, 
    ``c''), closed symbols denote macroscopic defects 1-3. 
    Solid lines indicate the lower and upper values,  $B_{FOT}^{min}$  
    and $B_{FOT}^{max}$, observed in this crystal. }
    \label{fig:crystal-1_5}
\end{figure}

DMO images on crystals from both boules reveal that the 
topology of the FOT process is dominated by crystalline 
inhomogeneity on the scale of $1-10$ $\mu$m. Figure \ref{fig:crystal-1_5} shows a 
series of four images of the FOT in crystal C (from boule 1),
measured at $T = 63.5$ K. The FOT in this and nearly all other investigated underdoped crystals
starts at inhomogeneities near the crystal edge. Unlike observations 
made on optimally doped crystals \cite{Soibel2000}, 
the FOT front always moves inwards from the edge towards the center. 
The distribution of local FOT fields has a temperature-independent 
width $\sigma_{FOT} = H_{FOT}^{max} - H_{FOT}^{min}$ of typically 5 Oe 
[Fig.~\ref{fig:crystal-1_5}(b)], twice that observed in optimally 
doped crystals \cite{Soibel2001}. The temperature range of the 
``collapse'' of $H_{p}$ and $H_{FOT}$ in the crystal center, 63 $< T <$ 65 K,
corresponds quite accurately to that over which $H_{p}$ attains 
$H_{FOT}^{min}$.

%% Fig 6 : Bean explanation
\begin{figure}[ht]

\centerline{\epsfxsize 12cm \epsfbox{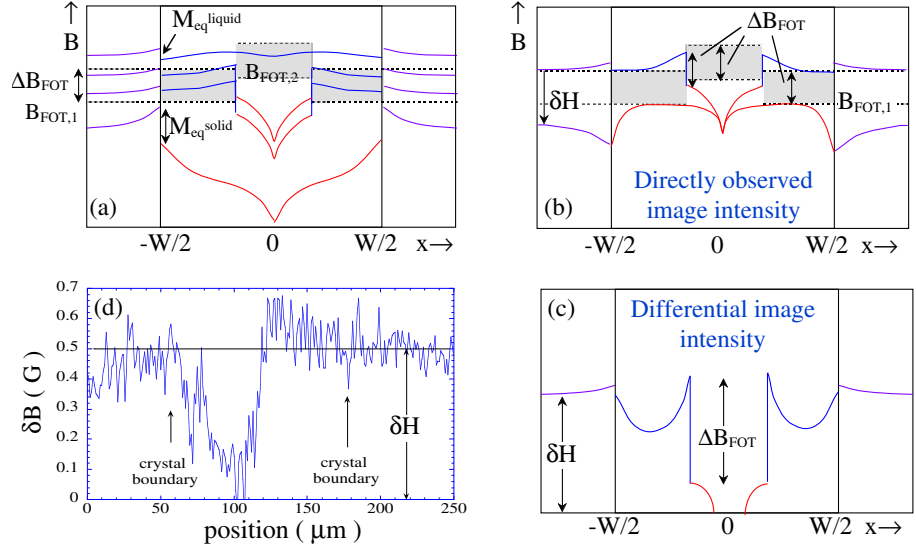}}
\vspace{-1cm}
\caption{Flux profiles in an inhomogeneous flat superconductor, containing 
a peripheral (1) and a central (2) region with FOT fields 
$B_{FOT,1} < B_{FOT,2}$. Hatched bars indicate the forbidden region
in the phase diagram, of width  $B_{FOT}$. Flux pinning 
is assumed to be present at $B < B_{FOT}$ and negligible for $B > B_{FOT}$.
(a) Increasing field.  (b) Profiles before and after a downward step $\delta 
H$ of the applied field, starting from the situation in which the FOT 
front has progressed to the boundary of region 1.
(3) Difference between the two profiles in (b). (d) 
Experimental profile across the DMO image at 1.5 
Oe in Fig.~\protect\ref{fig:crystal-1_5}(a). }
\label{fig:Bean}
\end{figure}	    

\section{Discussion and Conclusion}

A (sudden) decrease of the ``penetration field'' $H_{p}$ must be
due to the decrease of the surface or geometrical barrier limiting vortex entry 
into the sample \cite{Indenbom94,Zeldov94II}. The condition for vortex 
entry is that the 
(Meissner) screening current at the sample edge be equal to a 
characteristic current (typically the depairing current). This 
condition can be met more easily (and the barrier suppressed) in 
two ways: an increase of the local Meissner current by  
modification of the sample geometry, or the suppression of the 
characteristic current, {\em e.g.} by (the vortex lattice transition 
at) local inhomogeneities \cite{Noriko}. The DMO observations suggest
that the latter is at the origin of the $H_{p}$ ``collapse'' in 
single crystalline Bi-2212.

Once the barrier is overcome in the weakest spot of the crystal,
the very same crystalline inhomogeneity takes on another role: the internal 
barriers associated with local compositional variations  
constitute a mesoscopic pinning potential impeding vortex motion \cite{Clem73II}.
The phase transformation front thus moves in from the edge towards the 
center. 

To model the motion of the FOT front, we assume that the vortex 

\noindent solid phase has a non-zero critical current, while the (high field) 
vortex liquid has negligible critical current. Then, the flux distribution 
in a model inhomogeneous sample containing a peripheral (1)
and a central (2)  region with FOT fields $B_{FOT,1} < B_{FOT,2}$ 
respectively, should be as in Fig.~\ref{fig:Bean}(a). At 
low field, one has the usual flux distribution due to the critical 
state \cite{Brandt96}. As $B_{FOT,1}$ is passed, vortex liquid regions  
enter the crystal from the edge. As long as the local induction does 
not exceed $B_{FOT} + \Delta B_{FOT}$, this implies the 
presence of a phase mixture adjacent to the FOT transformation front. 
Within homogeneously disordered regions ({\em e.g.} 1 and 2), in which
the length scale characterising the defect structure is much smaller 
than the vortex spacing, the front motion is limited only by the 
dynamics of the liquid state and of the phase transformation itself. 
If the vortex liquid has negligible critical current, the front 
can move $freely$ through the crystal until it meets a region of higher
local $B_{FOT,2}$, eliminating local screening currents due to pinning 
in the process (a vortex solid-liquid intermediate state may be
left behind the FOT front). If, as in a DMO experiment, the applied field 
is lowered by $\delta H$, the FOT front recedes to the regions with 
lower local $B_{FOT}$, but non-equilibrium screening currents associated
with this retreat may be insufficient to change the local induction in the
resolidified region [Fig.~\ref{fig:Bean}(b)]. A differential image 
[Fig.~\ref{fig:Bean}(c)] reveals the extremal position of the FOT front 
as paramagnetic only because of the demagnetizing field around 
regions with high local $B_{FOT}$. These screen the 
modulation $\delta H$, and therefore show up as black, as in 
Figs.~\ref{fig:BendingRef} and \ref{fig:crystal-1_5}.
If $\delta H$ much exceeds $\Delta B_{FOT}$, {\em e.g.} 
near $T_{c}$ \cite{Zeldov95II}, the discontinuity in local B
is masked by the induced screening current in the regions 
with higher $B_{FOT}$; features associated with the FOT can
then no longer be observed (see {\em e.g.} the curve at 70 K in 
Fig.\ref{fig:Hall}).

Summarizing, the First Order Transition of the vortex lattice was
observed in underdoped Bi$_{2}$Ca$_{2}$CaCu$_{2}$O$_{8}$. The 
specific features related to the FOT in this material can be well 
understood as arising from mesocopic inhomogeneity. Our results 
constitute new proof of the robustness of the FOT to substantial amounts of 
crystalline disorder \cite{Banerjee2003}.

%The contrast in regions behind the FOT front is determined by the proportion 
%of solid and liquid phase. The grey level in 
%Fig.~\ref{fig:crystal-1_5}(a) suggests that for the $\delta H$ used 
%in our experiments, these regions consist (nearly) completely of flux 
%liquid. 

\end{document}